\documentclass[aip,jcp,amsmath,amssymb,reprint,superscriptaddress]{revtex4-2}
\usepackage[dvipdfmx]{graphicx}
\usepackage{url}
\usepackage{epsf}
\usepackage{bm}
\usepackage{lipsum}
\usepackage[normalem]{ulem}
\usepackage{txfonts}
\def\gs{\gtrsim}

\usepackage{color}

\usepackage{amsmath}	
\begin{document}
\title{Effects of chain length on Rouse modes and non-Gaussianity in
linear and ring polymer melts}

\author{Shota Goto}
\affiliation{Division of Chemical Engineering, Department of Materials Engineering Science, Graduate School of Engineering Science, Osaka University, Toyonaka, Osaka 560-8531, Japan}

\author{Kang Kim}
\email{kk@cheng.es.osaka-u.ac.jp}
\affiliation{Division of Chemical Engineering, Department of Materials Engineering Science, Graduate School of Engineering Science, Osaka University, Toyonaka, Osaka 560-8531, Japan}

\author{Nobuyuki Matubayasi}
\email{nobuyuki@cheng.es.osaka-u.ac.jp}
\affiliation{Division of Chemical Engineering, Department of Materials Engineering Science, Graduate School of Engineering Science, Osaka University, Toyonaka, Osaka 560-8531, Japan}

\date{\today}

\begin{abstract}
The dynamics of ring polymer melts are studied via molecular dynamics
 simulations of the Kremer--Grest bead-spring model.
Rouse mode analysis is performed
in comparison with linear polymers by changing the chain length.
Rouse-like behavior is observed in ring polymers by quantifying 
the chain length dependence of the Rouse relaxation time,
 whereas a crossover from Rouse to reptation behavior is
 observed in linear polymers.
Furthermore, the non-Gaussian parameters of the monomer bead
 displacement and chain center-of-mass displacement are analyzed.
It is found that the non-Gaussianity of ring polymers is remarkably
 suppressed with slight growth for the center-of-mass dynamics
 at long chain length, which is in contrast to the growth in
 linear polymers both for the monomer bead and center-of-mass dynamics.
\end{abstract}

\maketitle

\section{Introduction}

The dynamics of polymer melts are governed 
by topological constraints, and due to the constraints, the viscosity and
relaxation time increase drastically 
with increasing degree of polymerization.
Linear chain ends play a significant role in determining the
slip motion of a single polymer chain, which is characterized by
the well-established reptation model.~\cite{Doi:1986ut}
Recently, another type of topological constraints in polymer has been
proposed, namely, 
ring polymer melts without chain ends.~\cite{Cates:1986kt,
Obukhov:1994ep, Muller:1996ko, Muller:2000bc, McLeish:2002ie}

Various molecular dynamics (MD) simulations have been performed 
to elucidate the topological constraint effects in ring polymer
melts.~\cite{Brown:1998cn,
Jang:2003cs, Tsolou:2010kf, Halverson:2011jg, Halverson:2011ee,
Rosa:2014jr, Tsalikis:2016hy, Tsalikis:2017jc}
In this regard, 
the chain length $N$ dependence of dynamical properties is the central
topic.
Tsolou \textit{et al.} reported MD simulation results of a united-atom
model for ring polyethylene 
melts with $N$ ranging from 24 to 400.~\cite{Tsolou:2010kf}
They demonstrated that the Rouse model is approximately appropriate for describing
the dynamics, in contrast to the cases of linear polymer analogues.
Halverson \textit{et al.} used a coarse-grained bead-spring model for ring
polymers with $N$ ranging from 100 to 1600.~\cite{Halverson:2011jg, Halverson:2011ee}
The diffusion coefficient $D$ obeys a scaling $D\sim N^{-2.4}$ for large
$N$--interestingly, this is 
 similar to that observed in linear polymer melts.
In contrast, the zero-shear viscosity exhibits a chain length
dependence $\eta\sim N^{1.4}$, which is weaker than that predicted by the
reptation model.

The dynamics of ring polymer melts have been examined using 
the dynamic structure factor measured by neutron scattering
experiments.~\cite{Bras:2011fz, Bras:2014ba, Goossen:2014eu, Kruteva:2020bt, Arrighi:2020gx}
Br\'{a}s \textit{et al.} reported the non-Gaussian parameter (NGP) of pure
poly(ethylene oxide) (PEO) rings.~\cite{Bras:2014ba}
The NGP characterizes the degree of 
the deviation 
of the distribution function of the monomer displacement 
 from the Gaussian distribution, which is important
when discussing the relationship between MD simulations
and scattering experiments.~\cite{Arbe:2012jh}
Notably, the NGP has frequently been analyzed to characterize 
heterogeneous dynamics, which is attributed to cage effects in
glass-forming liquids.~\cite{Kob:1997hl, Donati:1999cn, Saltzman:2006kj}
However, the chain length dependence of NGP in ring polymers remains
scarcely analyzed.
Furthermore, this analysis can be also important when considering the recent microscopic
theory predicting $D\sim N^{-2}$ in ring polymer melts, which was
formulated in analogy
with the cage effects of soft colloid suspensions.~\cite{Mei:2020ji}

In this study, 
we performed MD simulations using the Kremer--Grest bead-spring
model with different chain lengths ($N=5-400$) for both linear and ring
polymer melts.
First, we analyzed the Rouse modes and determined the chain length
dependence of the relaxation time.
Then, we calculated the NGP of the monomer bead displacement, and
investigated its chain length dependence.
The combined results enable us to thoroughly assess the 
similarities and differences of the chain-end effects 
on the dynamics between linear and ring polymer melts.

\section{Model and simulations}

We performed MD simulations using the standard Kremer--Grest model for linear and ring polymer melts, where
the polymer chain comprises $N$ monomer beads of mass $m$ and
diameter $\sigma$.~\cite{Kremer:1990iv}
We utilized three types of inter-particle potentials, as follows.
The Lennard-Jones (LJ) potential
\begin{equation}
U_\mathrm{LJ}(r) = 4\varepsilon_\mathrm{LJ} \left[ \left(
						    \frac{\sigma}{r}
						   \right)^{12} - \left(
						   \frac{\sigma}{r}
								  \right)^{6}
					   \right] +C,
\label{eq:LJ}
\end{equation}
acts between all pairs of monomer beads, where $r$ and
$\varepsilon_\mathrm{LJ}$ denote the distance between two monomers and
the energy scale of the LJ potential, respectively.
The LJ potential is truncated at the cut-off distance of $r_c
=2.5\sigma$, and the constant $C$ guarantees that the potential energy
shifts to zero at $r=r_c$
The bonding potential between two neighboring monomer beads is given 
by a finitely extensible nonlinear elastic (FENE) potential,
\begin{equation}
U_\mathrm{FENE}(r) =
-\frac{1}{2} K R_0^2 \ln \left[ 1 -
 \left(\frac{r}{R_0} \right)^{2} \right]
\label{eq:FENE}
\end{equation}
for $r < R_0$, where $K$ and $R_0$ represent the spring
constant and the maximum length of the FENE bond, respectively.
We used the values of $K = 30\varepsilon_\mathrm{LJ}/\sigma^2$ and $R_0 = 1.5\sigma$.
Finally, the bending angle $\theta$ formed by three consecutive monomer
beads along the polymer chain is controlled by
\begin{equation}
U_\mathrm{bend}(\theta) = k_\theta \left[ 1 - \cos(\theta - \theta_0) \right],
\label{eq:bend}
\end{equation}
where $k_\theta$ denotes the associated bending energy.
We set the bending energy and equilibrium angle as
$k_\theta=1.5\varepsilon_\mathrm{LJ}$ and $\theta_0 = 180^\circ$, respectively.

Henceforth, the length, energy, and time are measured in units of
$\sigma$, $\varepsilon_\mathrm{LJ}$, and
$\sigma(m/\varepsilon_\mathrm{LJ})^{1/2}$, respectively.
The temperature is presented in units of
$\varepsilon_\mathrm{LJ}/k_\mathrm{B}$, where $k_\mathrm{B}$ is the
Boltzmann constant. 
The system contains $M$ polymer chains in a
three-dimensional cubic box of volume $V$ under periodic boundary
conditions.
We studied several combinations of the chain length $N$ and the number of
chains $M$ for both linear and ring polymer systems, 
 $(N, M)=(5, 2000)$, $(10, 1000)$, $(20, 500)$, $(40, 250)$, $(100,
 200)$, $(200, 100)$, and $(400, 50)$.
The number density of the monomer beads
$\rho=(N\times M)/V$ and the temperature $T$ were fixed as 
$\rho=0.85$ and $T=1.0$, respectively, throughout the simulations.
We performed the MD simulations using the Large-scale Atomic/Molecular
Massively Parallel Simulator (LAMMPS).~\cite{Plimpton:1995wl}
The $NVT$ ensemble with the Nos\'{e}--Hoover thermostat was used with a time step $\Delta t$ of 0.01.
We analyzed the chain length dependence of the radius of
gyration and the center-of-mass diffusion coefficient and confirmed
that our results reproduce the results reported in previous
studies (results not shown).~\cite{Halverson:2011jg, Halverson:2011ee}
In addition, we confirmed the entanglement length
$N_\mathrm{e}\approx 28$ in linear
polymer melts with $N=400$ by using 
the primitive path analysis.~\cite{Sukumaran:2005gy}
The used code is available from
\mbox{\url{https://github.com/t-murash/USER-PPA}} (see also
Ref.~\onlinecite{Hagita:2021jk}).


The Rouse model is the standard model for the polymer chain dynamics, where 
the normal coordinates $\bm{X}_p(t)$, so-called Rouse modes, are constructed from the position of
the $n$-th monomer bead $\bm{r}_n(t)$ at a time $t$ for $n=1$, 2, 3, $\cdots$, $N$.
Here, we provide several expressions in the Rouse model, which we employ to analyze our MD results.
The Rouse mode analysis for the linear chain is described in Ref.~\onlinecite{Kopf:1997ht}.
Furthermore, the formula for the ring polymer chain was 
described in previous papers.~\cite{Wiest:1987bn, Tsolou:2010kf, Rauscher:2020jz}
To make this paper self-contained, 
we summarize the formulation of the Rouse model for the ring polymer chain in Appendix.
The expressions of the normal coordinates $\bm{X}_{p,
\mathrm{linear}}(t)$ and $\bm{X}_{p, \mathrm{ring}}(t)$
for linear and ring polymer chains can respectively be expressed as
\begin{align}
\bm{X}_{p, \mathrm{linear}}(t) &= \sqrt{\frac{2-\delta_{p,0}}{N}}\sum_{n=1}^N
 \bm{r}_n(t)\cos\left(\frac{\pi p(n-1/2)}{N}\right),
\label{eq:Rouse_mode_linear}
\\
\bm{X}_{p, \mathrm{ring}}(t) 
&=
 \sqrt{\frac{1}{N}}\sum_{n=1}^{N}
\bm{r}_n(t)\left[\cos\left(\frac{2\pi pn}{N}\right) + \sin\left(\frac{2\pi pn}{N}\right)\right],
\label{eq:Rouse_mode_ring}
\end{align}
where $p$ $(=0,1,\cdots, N-1)$ is the mode index, and $\delta$ denotes the Kronecker delta.
The $p=0$ mode describes the center-of-mass translation of the
chain, whereas the $p>0$ modes characterize the internal dynamics of the
subchains composed of $N/p$ beads.

The static correlation of the Rouse mode $\langle \bm{X}_p(0)^2\rangle$
can be related to the mean square
distance of two beads $b^2$ through
\begin{align}
\langle \bm{X}_{p, \mathrm{linear}}(0)^2\rangle &= \frac{b^2}{4\sin^2\left(\frac{\pi
 p}{2N}\right)},\label{eq:static_linear}\\
\langle \bm{X}_{p, \mathrm{ring}}(0)^2\rangle &= \frac{b^2}{4\sin^2\left(\frac{\pi
 p}{N}\right)},\label{eq:static_ring}
\end{align}
for linear and ring polymers, respectively.
Here, $\langle \cdots\rangle$ denotes an ensemble average.

Each normal coordinate exhibits the Brownian motion in the Rouse model, 
causing the exponential decay of the autocorrelation function, $\langle \bm{X}_p(t)\cdot\bm{X}_p(0)\rangle$.
The Rouse relaxation times $\tau_{p, \mathrm{linear}}$ and $\tau_{p, \mathrm{ring}}$
for linear and ring polymer chains are respectively given by 
\begin{align}
\tau_{p, \mathrm{linear}} &= \frac{\zeta}{4k\sin^2\left(\frac{\pi
 p}{2N}\right)},\label{eq:tau_p_linear}\\
\tau_{p, \mathrm{ring}} &= \frac{\zeta}{4k\sin^2\left(\frac{\pi p}{N}\right)}\label{eq:tau_p_ring},
\end{align}
where $\zeta$ is the effective hydrodynamic friction coefficient and 
$k$ represents the harmonic spring constant between two neighboring monomer
beads.
As noted in Appendix $k$ is equal to $3k_\mathrm{B}T/b^2$.
The differences of $\langle \bm{X}_p(0)^2\rangle$ and $\tau_p$ between
linear and ring polymers appear in the phases of the sine functions.
The Rouse modes of $p$ and $N-p$ are degenerate in
the case of the ring polymer (see Appendix).
Correspondingly, $\langle \bm{X}_p(0)^2\rangle$ and 
$\tau_p$ as functions of $p$ are symmetric with respect to the
reflection at $p=N/2$.
On the other hand, for linear chains, $\langle \bm{X}_p(0)^2\rangle$ and
$\tau_p$ decrease monotonically with $p$ in the Rouse model.
In the continuum limit of $p/N\ll 1$, both $\tau_{p, \mathrm{linear}}$ and
$\tau_{p, \mathrm{ring}}$ exhibit a scaling behavior $(N/p)^2$ within the
Rouse model.

\begin{figure*}[t]
\centering
\includegraphics[width=0.8\textwidth]{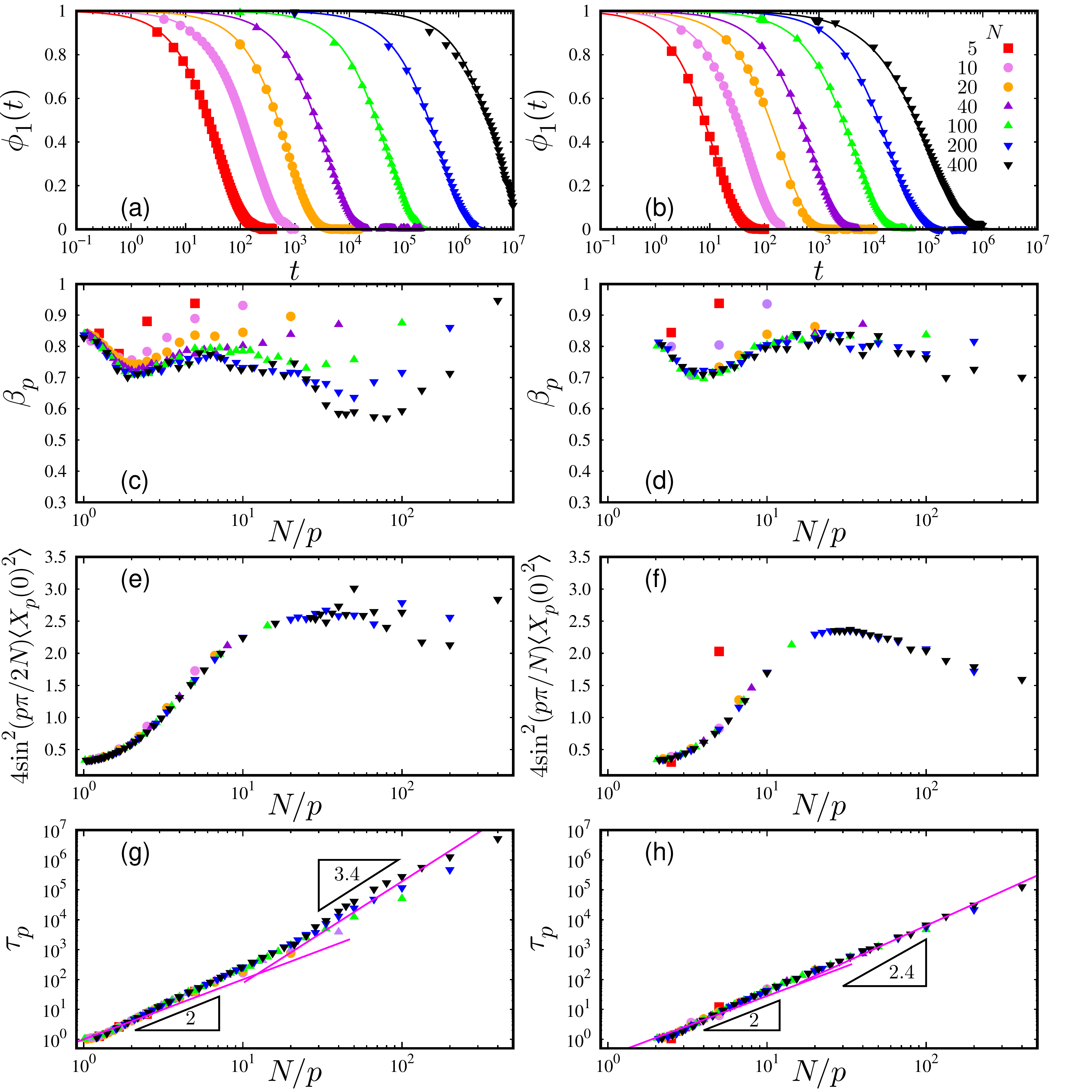}
\caption{
Normalized autocorrelation function 
 $\phi_1(t)$ of the Rouse mode $p=1$ for linear (a) and ring (b) polymers.
Symbols and lines represent MD simulation results and the fitting curves
 using the Kohlrausch--Williams--Watts function, $\exp[-(t/\tau_p)^{\beta_p}]$.
The exponent $\beta_p$ is plotted as a function of $N/p$ for linear (c)
 and ring (d) polymers.
Normalized amplitude of autocorrelations of the Rouse mode,
 $4 \sin^2(\pi p/(2N))\langle
 \bm{X}_p(0)^2\rangle$ (e) and $4 \sin^2(\pi
 p/N)\langle\bm{X}_p(0)^2 \rangle$ (f), are plotted as a function of
 $N/p$ for linear and ring polymers, respectively.
Rouse relaxation time $\tau_p$ as a function of $N/p$ for
 linear (g) and ring (h) polymers.
Two scaling behaviors, \textit{i.e.}, the Rouse model behavior 
 $\tau_p\sim (N/p)^{2}$ and the reptation model behavior $\tau_p\sim (N/p)^{3.4}$, are represented in (g).
In (h), $\tau_p \sim (N/p)^{2}$ is indicated for
 smaller $N/p$, whereas the different power-law $\tau_p\sim (N/p)^{2.4}$
 is observed for larger $N/p$.
In (d), (f), and (h), the results for $N/p < 2$ are omitted because of the symmetric
 structure of $N/p$ dependencies on $\langle \bm{X}_p(0)^2\rangle$ and
 $\tau_p$ (see Eqs.~(\ref{eq:Rouse_mode_ring}), (\ref{eq:static_ring}) and (\ref{eq:tau_p_ring})).
Note that only two points with $p=1$ and 2 are plotted for $N=5$
 ring polymers, where each ring tends to form a pentagonal structure,
 causing fluctuations more than other length chains.
}
\label{fig:Rouse}
\end{figure*}


The motions of monomer beads are described typically by the mean square
displacement (MSD) averaged over all the monomers of a chain, which is defined as
\begin{align}
g_1(t)= \langle r^2(t)\rangle = \left\langle \frac{1}{N}\sum_{n=1}^N|\bm{r}_n(t)-\bm{r}_n(0)|^2\right\rangle.
\label{eq:MSD}
\end{align}
The NGP of the monomer bead displacement is defined by 
\begin{align}
\alpha^\mathrm{mon}_2(t)  = \frac{3\langle r^4(t)\rangle}{5\langle r^2(t)\rangle^2}-1,
\label{eq:NGP_mon}
\end{align}
which measures non-Gaussianity, \textit{i.e.}, the degree of the deviation
of the distribution function of the monomer bead displacement from the
Gaussian form during the time interval $t$.
In addition, the MSD of the center-of-mass of chains is examined from
\begin{align}
g_3(t)= \langle R^2(t)\rangle = \left\langle
 \frac{1}{M}\sum_{m=1}^M|\bm{R}_m(t)-\bm{R}_m(0)|^2\right\rangle,
\end{align}
where $\bm{R}_m(t)$ is the position of
the center-of-mass of the chain $m$ at time $t$.
The corresponding NGP of the center-of-mass displacement is defined by 
\begin{align}
\alpha^\mathrm{com}_2(t)  = \frac{3\langle R^4(t)\rangle}{5\langle R^2(t)\rangle^2}-1.
\label{eq:NGP_com}
\end{align}
The NGP of monomer beads was analyzed via MD simulations of linear polymer melts with the
chain length of $N=5-160$.~\cite{Pan:2018ib}
Furthermore, the NGP of supercooled
polymer melts was reported with $N=10$~\cite{Aichele:2003dk} and $N=64$.~\cite{Peter:2009iq, Barrat:2010ce}

\section{Results and discussion}

\begin{figure}[t]
\centering
\includegraphics[width=0.4\textwidth]{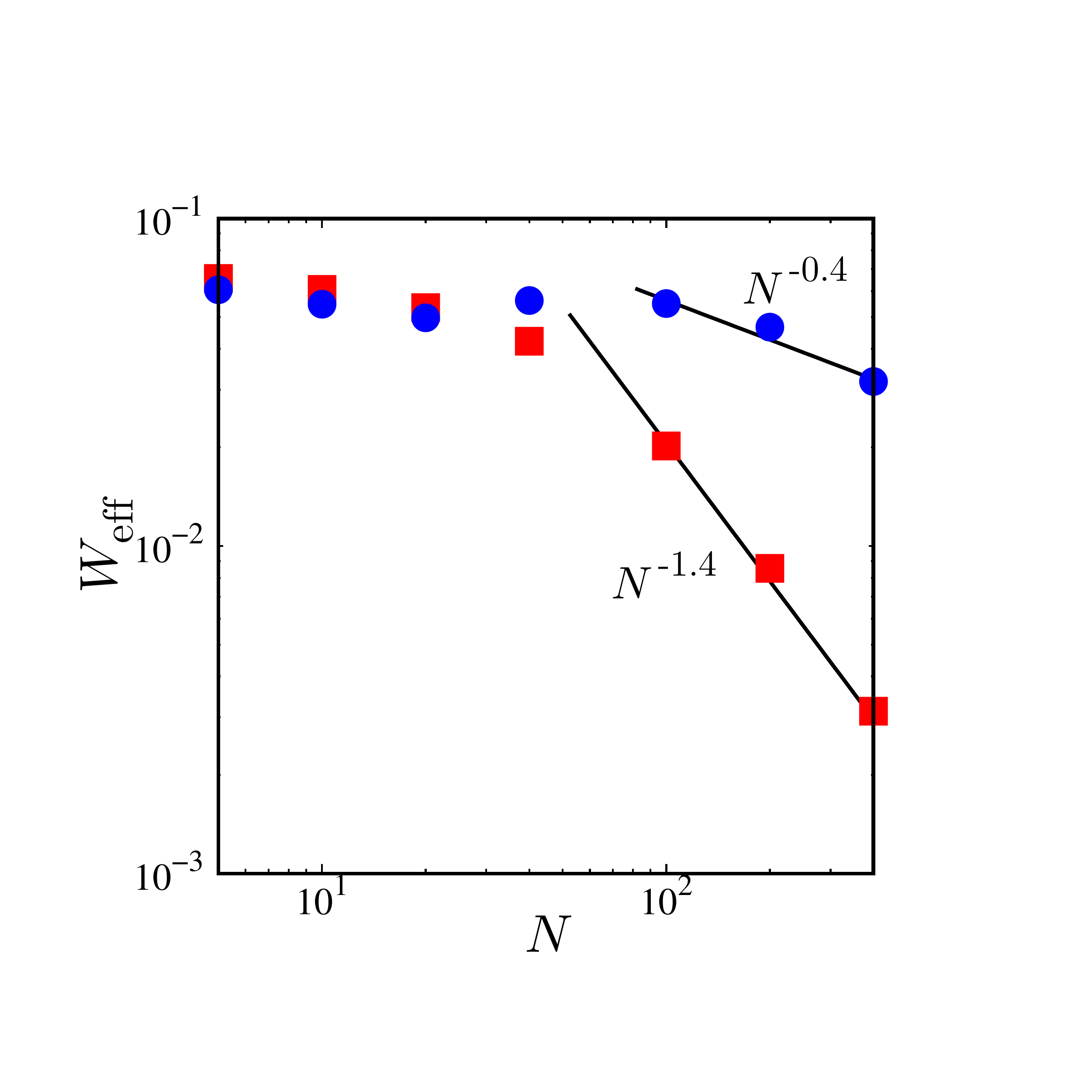}
\caption{
Chain length $N$ dependence of the effective segmental relaxation rate
 $W_\mathrm{eff}$ for linear (squares) and ring (circles) polymers.
Two straight lines are eye guides indicating, $W_\mathrm{eff} \sim N^{-1.4}$ and $W_\mathrm{eff} \sim N^{-0.4}$.}
\label{fig:W}
\end{figure}

The normalized autocorrelation function of the $p$-th Rouse mode is 
given by 
\begin{equation}
\phi_p(t) =\frac{\langle \bm{X}_p(t)\cdot \bm{X}_p(0)\rangle}{\langle
 \bm{X}_p(0)^2\rangle}.
\end{equation}
The results of the slowest mode $\phi_1(t)$ are plotted in
Fig.~\ref{fig:Rouse} by changing the chain length $N$ for linear (a) and ring (b) polymer melts.
For each Rouse mode $p$, 
$\phi_p(t)$ is fitted using the
Kohlrausch--Williams--Watts (KWW) function, $\exp[-(t/\tau_p^*)^{\beta_p}]$,
with the KWW relaxation time $\tau_p^*$.
$\beta_p(<1)$ represents the degree of non-exponentiality of $\phi_p(t)$.~\cite{Shaffer:1995bk}
In Fig.~\ref{fig:Rouse}(c) and (d), $\beta_p$ is plotted as a function
of $N/p$ for linear and
ring polymers, respectively.
As demonstrated in Ref.~\onlinecite{Padding:2002hd}, $\beta_p$ deviates
from unity and shows a minimum at the slowing down length $N_\mathrm{s}\approx 2$.
Another minimum of approximate 0.6 is found at around 
the entanglement length scale $N_\mathrm{e}\approx 28$.~\cite{Kalathi:2014iq, Hsu:2017cr}
As seen in Fig.~\ref{fig:Rouse}(d), $\beta_p$ of ring polymers shows a minimum at $N_\mathrm{s}\approx 4$,
which is the same length scale of $N_\mathrm{s}\approx 2$
considering the difference in the phase of the Rouse mode between linear and
ring polymers.
Furthermore, the non-exponentiality is also found at $N \gtrsim 10^{2}$
and is weaker for the ring polymers with $\beta_{p} \approx 0.8$
than for the linear polymers.

In Fig.~\ref{fig:Rouse}(e) and (f), the normalized amplitudes
$4\sin^2(\pi p/(2N))\langle \bm{X}_p(0)^2\rangle$ and $4\sin^2(\pi
p/N)\langle \bm{X}_p(0)^2\rangle$
are plotted as a function of $N/p$ for linear and
ring polymers, respectively.
As the chain length scale $N/p$ increases, 
$4\sin^2(\pi p/(2N))\langle
\bm{X}_p(0)^2\rangle$ of linear polymers levels
off beyond the entanglement length scale $N_\mathrm{e}\approx 28$~\cite{Kalathi:2014iq, Hsu:2017cr}, 
whereas $4\sin^2(\pi p/N)\langle \bm{X}_p(0)^2\rangle$ of ring polymers gradually
decreases with increasing $N/p$.
This behavior is actually consistent with the observation that the
structure of the ring polymer chain becomes more compact than that of 
the linear polymer.
In fact, $N$ dependence of 
the mean square radius of gyration $R_\mathrm{g}^2$ approaches a scaling of
$N^{2/3}$ in ring polymers, which is distinct from the Gaussian
behavior $R_\mathrm{g}^2 \sim N$ observed in linear polymers.~\cite{Halverson:2011jg}

The effective Rouse relaxation time of the $p$-th mode is calculated by
\begin{equation}
\tau_p = \int_0^\infty \exp[-(t/\tau_p^*)^{\beta_p}] dt =
 \frac{\tau_p^*}{\beta_p}\Gamma\left(\frac{1}{\beta_p}\right),
\end{equation}
where $\Gamma(x)$ is the Gamma function.
The Rouse relaxation time $\tau_p$ is plotted as a function of $N/p$ in 
Fig.~\ref{fig:Rouse} for linear (g) and ring (h) polymer melts.
In linear polymer melts, 
$\tau_p$ rapidly deviates from the Rouse regime $(N/p)^2$ 
as 
the chain length $N$ is increases.
In particular, the power-law behavior $\tau_p\sim (N/p)^{3.4}$ was
observed, indicating entanglement effects.~\cite{Kalathi:2014iq, Hsu:2017cr}
This crossover from the Rouse to the
reptation behavior was reported in Refs.~\onlinecite{Kalathi:2014iq, Hsu:2017cr}.
$\tau_p$ of ring polymers also deviates from the Rouse-like power-law
behavior with increasing $N/p$.
However, the exponent becomes 2.4, which is smaller than that of 
linear polymers for the chain lengths investigated in this study.

\begin{figure}[t]
\centering
\includegraphics[width=0.5\textwidth]{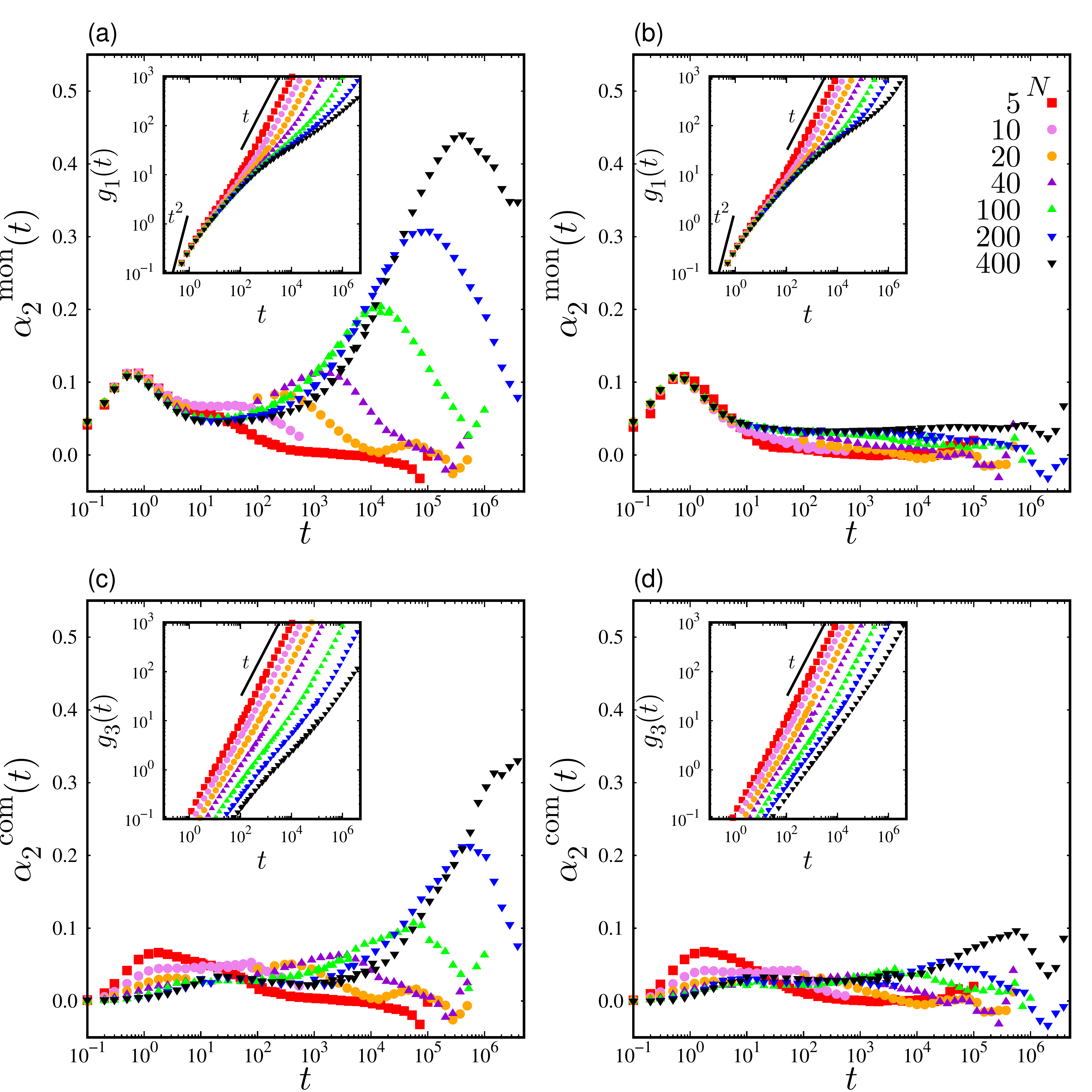}
\caption{Non-Gaussian parameters of monomer beads $\alpha^\mathrm{mon}_2(t)$ and
 center-of-mass $\alpha^\mathrm{com}_2(t)$
 for linear [(a) and 
 (c)] and ring [(b) and (d)] polymers.
Inset: Mean square displacements of monomer beads $\langle r^2(t)\rangle$ 
 and the center-of-mass $\langle R^2(t)\rangle$.
The solid lines represent ballistic motion $(\sim t^2)$
 and diffusive behavior $(\sim t)$.
}
\label{fig:NGP}
\end{figure}

Further, it is important to compare the segmental relaxation rate
$W_\mathrm{eff}=3k_\mathrm{B}T/\zeta b^2 = k/\zeta$ between linear and ring polymer melts, which
is related to the Rouse relaxation time $\tau_{p}$ (see Eqs.~(\ref{eq:tau_p_linear}) and (\ref{eq:tau_p_ring})).
Specifically, we evaluated 
$W_\mathrm{eff}$ using the slowest mode $(p=1)$ by 
\begin{align}
W_\mathrm{eff,linear} &= 1/[4\tau_\mathrm{1,linear} \sin^2(\pi /2N)],\\
W_\mathrm{eff,ring} &= 1/[4\tau_\mathrm{1,ring} \sin^2(\pi /N)],
\end{align}
for linear and ring polymers, respectively, and the results are plotted in Fig.~\ref{fig:W}.
For linear
polymers, $W_\mathrm{eff}$ exhibits a roughly constant independent
of $N$ up to the entanglement length $N_\mathrm{e}\approx 28$.
A similar value is also observed for ring polymers, indicating the
same Rouse dynamics in melts of linear and ring chains.
The power-law behavior $W_\mathrm{eff}\sim N^{-1.4}$ is observed for the
longer linear polymer, which is consistent with the scaling of
$\tau_p\sim (N/p)^{3.4}$, as demonstrated in Fig.~\ref{fig:Rouse}(e).
Note that $N$ and $p$ are both varied in Fig.~\ref{fig:Rouse}(e), and the
scaling at $p =1$ is rephrased as $\tau_1 \sim N^{3.4}$ at large $N$.
In contrast, 
$W_\mathrm{eff}$ of ring polymers shows a weak $N$ dependence and the
scaling $W_\mathrm{eff}\sim N^{-0.4}$ is observed for the longer chain
length $N\gs 100$.
This exponent corresponds to the scaling of $\tau_p\sim
(N/p)^{2.4}$, as observed in Fig.~\ref{fig:Rouse}(f).

\begin{figure}[t]
\centering
\includegraphics[width=0.5\textwidth]{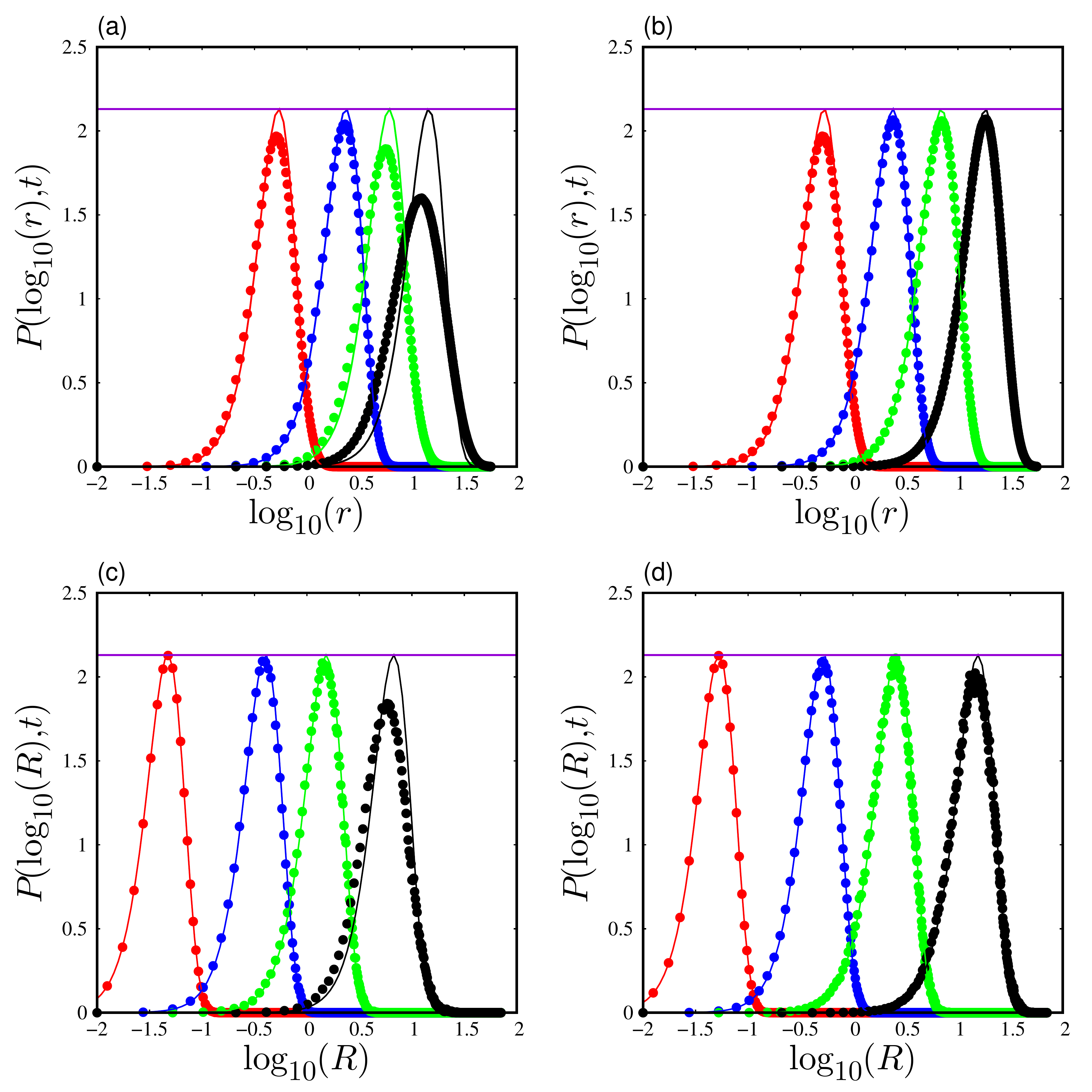}
\caption{Probability distributions of the logarithm displacement of
 monomer beads and center-of-mass, 
 $P(\log_{10}(r), t)$ and $P(\log_{10}(R), t)$, for linear [(a) and (c)] and ring
 [(b) and (d)] polymers with the chain length $N=400$.
The time $t$ is chosen as $t=1$, $10^2$, $10^4$, and $10^6$ from left to right.
The horizontal line denotes the Gaussian level, $\ln(10)\sqrt{54/\pi}e^{-3/2}\approx 2.13$.
The solid curve represent the form using the Gaussian distribution,
$G_s(r, t)=[3/(2\pi \langle r^2(t)\rangle)]^{3/2}\exp[-3r^2/(2\langle
 r^2(t)\rangle)]$ [(a) and (b)] and $G_s(R, t)=[3/(2\pi \langle R^2(t)\rangle)]^{3/2}\exp[-3R^2/(2\langle
 R^2(t)\rangle)]$ [(c) and (d)] at each time.
}
\label{fig:van_Hove}
\end{figure}

The NGPs of the segment displacement $\alpha^\mathrm{mon}_2(t)$ and 
the center-of-mass displacement $\alpha^\mathrm{com}_2(t)$ were
investigated using Eqs.~(\ref{eq:NGP_mon}) and (\ref{eq:NGP_com}), respectively.
Figure~\ref{fig:NGP} shows $\alpha^\mathrm{mon}_2(t)$ for linear (a) and ring (b) polymers.
For comparison, the time evolusions of MSD $\langle r^2(t)\rangle$ are displayed
in inset of Fig.~\ref{fig:NGP}(a) and (b).
It is seen that $\alpha^\mathrm{mon}_2(t)$ exhibits peaks of 0.1 for both linear and ring
polymers.
The peak occurs at $t \approx 1$, beyond which each segment begins to
escape from the regime of ballistic motion, $\langle r^2(t)\rangle \sim t^2$, at
small times.
The height and position $\alpha^\mathrm{mon}_2(t)$ in the ballistic regime are independent of the chain
length $N$, indicating that the effects of polymer chain ends are
negligible in this regime, where the effect of the chain connectivity
plays the role on the segmental dynamics.~\cite{Aichele:2003dk}
For linear polymers, the second peak appears at a larger time
regime, where $\langle r^2(t)\rangle$ approaches the diffusive behavior,
as demonstrated in Fig.~\ref{fig:NGP}(c).
The second peak develops for longer time
scales with increasing chain length $N$, which was demonstrated in the previous study.~\cite{Pan:2018ib}
The height of the second peak becomes 0.5 for $N=400$.
This non-Gaussianity can be regarded as the chain end effect with higher
mobility due to less topological constrains.~\cite{Wang:2012fy}
Note that the mechanism of non-Gaussianity in linear polymer melts is different from
that of the cage effects in glass-forming liquids.~\cite{Kob:1997hl,
Donati:1999cn, Saltzman:2006kj}
On the contrary, it is unlikely that $\alpha^\mathrm{mon}_2(t)$ of ring polymers
shows clear peaks for chain lengths up to $N=400$ despite the diffusive behavior being realized in
$\langle r^2(t)\rangle$ at larger time scales (see Fig.~\ref{fig:NGP}(d)).
This implies that all the monomer beads show similar dynamics in ring
polymers without chain ends.
The non-Gaussianity of the center-of-mass displacement is additionally 
examined in Fig.~\ref{fig:NGP} for linear (c) and ring (d) polymers.
The behavior of $\alpha^\mathrm{com}_2(t)$ is analogous to that of
$\alpha^\mathrm{mon}_2(t)$ both for linear and ring polymers.
However, the first peak
of $\alpha^\mathrm{com}_2(t)$ at $t\approx 1$ becomes smaller with
increasing $N$ for both linear and ring polymers.
This indicates that regardless of the chain connectivity, the center-of-mass
dynamics is more Gaussian for longer chains.
Furthermore, $\alpha^\mathrm{com}_2(t)$ of ring
polymers with $N=400$ shows a peak of 0.1 at $t\approx 10^6$, which
shows very small non-Gaussianity as the chain length is increased.

Finally, to characterize the difference in the NGP between linear and ring polymers in more detail, 
we calculated the self-part of the van Hove correlation function $G_s(r,
t)= \langle \sum_{n=1}^N\delta(|\bm{r}_n(t)-\bm{r}_n(0)|-r)\rangle$,
\textit{i.e.}, the distribution function of the segmental displacement
$r$ at time $t$.
The probability distribution of the logarithm displacement is then 
defined as $P(\log_{10}(r), t) = \ln (10) 4\pi r^3 G_s(r,
t)$.~\cite{Cates:2004cn, Reichman:2005eq, Flenner:2005es}
It is defined such that 
the integral $\int_{x_0}^{x_1}P(x, t)dx$ is the fraction of particles whose value of
$\log_{10}(r)$ is between $x_0$ and $x_1$.
When the Gaussian distribution is assumed as $G_s(r, t)=[3/(2\pi \langle
r^2(t)\rangle)]^{3/2}\exp[-3r^2/(2\langle
r^2(t)\rangle)]$ with the mean square displacement $\langle
r^2(t)\rangle$ at time $t$, 
$P(\log_{10}(r), t)$ has a peak of $\ln(10)\sqrt{54/\pi}e^{-3/2}\approx
2.13$ irrespective of time $t$.
In Fig.~\ref{fig:van_Hove}, $P(\log_{10}(r), t)$ is plotted for linear (a)
and ring (b) polymers with the chain length $N=400$ by changing $t$ from 1 to $10^6$.
For a comparison, we also showed $P(\log(10)(r), t)$
determined from the Gaussian distribution $G_s(r, t)=[3/(2\pi \langle
r^2(t)\rangle)]^{3/2}\exp[-3r^2/(2\langle
r^2(t)\rangle)]$ at each time.
As observed in Fig.~\ref{fig:van_Hove}(a), 
the peak height of $P(\log_{10}(r), t)$ for the
linear polymer decreases as $t$ increases.
This decrease in the peak indicates that the distribution deviates
from the Gaussian behavior and becomes broader, which is also observed
in glass-forming liquids.~\cite{Flenner:2005es}
Figure~\ref{fig:van_Hove}(b) demonstrates that the peak height
of $P(\log_{10}(r), t)$ for ring polymers remains at the Gaussian level, 
providing clear evidence that the segment
displacement follows the Gaussian distribution even for longer time scales.

Furthermore, Fig.~\ref{fig:van_Hove}(c) and
Fig.~\ref{fig:van_Hove}(d) show the probability distributions of the
center-of-mass displacement $P(\log_{10}(R), t)$ for linear and ring
polymers, respectively.
The deviation from the Gaussian form $G_s(R, t)=[3/(2\pi \langle
R^2(t)\rangle)]^{3/2}\exp[-3R^2/(2\langle
R^2(t)\rangle)]$ is noticeable for linear polymers, particularly for
longer times.
Analogous to Fig.~\ref{fig:van_Hove}(b), $P(\log_{10}(R), t)$ of ring
polymers is in accordance with the Gaussian distribution at any time.
Note that 
small deviation from the Gaussian distribution at long times were observed 
in Fig.~\ref{fig:NGP}(d) for the chain length $N=400$, 
while the peak value remains the Gaussian
level of 2.13.
This 
observation concerning the loss of Gaussianity suggests the possibility
that the center-of-mass dynamics of a long ring
polymer chain in melts can be influenced by the neighboring rings.

\section{Conclusions and final remarks}

We presented the MD simulation results using the
Kremer--Grest model for linear and ring polymer melts with
chain lengths up to $N=400$.
We focused on the chain length dependence of
the Rouse relaxation time and
non-Gaussianity for characterizing both the segmental and center-of-mass
mobility with or
without chain ends.

For linear polymers, the deviation from the Rouse model behavior
becomes remarkable with
increasing the chain length $N$ by showing the scaling $\tau_p \sim
(N/p)^{3.4}$, which is consistent with previously reported results.~\cite{Halverson:2011ee}
The NGP of the monomer bead dynamics shows two peaks: the first peak appears on the
time scale where the MSD escapes from the segmental ballistic motion, whereas 
the second peak corresponds to the realization of the
diffusive behavior of the MSD.
This indicates that the segment dynamics becomes spatially
heterogeneous because of the higher mobility of chain ends in
the linear polymer chain.
The NGP of the center-of-mass dynamics also exhibits two peaks, but the
first peak becomes weaker due to less
chain connectivity effects as the chain length is increased.

For ring polymers, the Rouse-like behavior with the scaling $\tau_p\sim
(N/p)^{2.4}$ was observed.
Although the peak of NGP was observed at short times similar to that of linear
polymers, 
the non-Gaussianity was found to be strongly
suppressed even for a longer time regime.
The segmental dynamics in ring polymers without
chain ends becomes spatially homogeneous and the mechanism of the
chain motion is essentially different from the reptation model for 
linear polymers.
The center-of-mass dynamics in ring polymers also shows the
Gaussian behavior, while
a very small non-Gaussianity is observed with increasing chain length
suggesting cooperative motions between neighboring rings.

As mentioned in Introduction, 
Br\'{a}s \textit{et al}. reported the NGP of the center-of-mass dynamics in PEO ring polymers from a neutron scattering
experiment.~\cite{Bras:2014ba}
The molecular weight 5 kg/mol was chosen to be 2.5 times larger than the
entanglement mass, which approximately corresponds to the chain length
$N=100$ in the present MD simulation study.
The NGP from the neutron scattering experiment
shows a peak of 0.2-0.3 at around 30 ns, which
corresponds to the crossover from a sub-diffusion to diffusion regime.
It seems that the experimental result is not in agreement with the present MD
simulation result of $\alpha^\mathrm{com}_2\approx 0.1$ with $N=400$.
The effects of the chain lengths and the chemical species of the segments 
need to be studied in further depths to resolve the difference.

A plausible key feature for topological constraints in ring polymers is an inter-ring
threading event.~\cite{Michieletto:2014ep, Michieletto:2014ee, Lee:2015dx, Michieletto:2016es,
Michieletto:2017fc, Michieletto:2017gh, Sakaue:2018jm, Lee:2019hq, Michieletto:2021hq}
In particular, Michieletto \textit{et al.} have proposed the ``random
pinning'' procedure, wherein some fractions of rings are
frozen, to investigate the role of threadings on the dynamics.~\cite{Michieletto:2017gh}
They demonstrated that random pinning can enhance the glass-like heterogeneous dynamics in ring polymers.
Furthermore, it was reported that the distribution of the center-of-mass
displacement deviates from the Gaussian distribution even in a zero
``random pinning'' field. 
In contrast, the non-Gaussianity is much weaker in this work,
where $\alpha^\mathrm{com}_2(t)$ becomes 0.1 with the chain length
$N=400$ without the pinning procedure.
One possible interpretation could be that the thermodynamic states
analyzed here 
are different: monomer density $\rho=0.85$ in this study is
frequently used for MD simulations of polymer
melts~\cite{Halverson:2011jg, Halverson:2011ee}, whereas densities in
Ref.~\onlinecite{Michieletto:2017gh} were chosen up to $\rho=0.4$.
Therefore, further investigation is necessary for a strict assessment with
regard to the monomer density dependence of 
the non-Gaussianity with increasing the chain length $N$, which is
a subject of future study.

\appendix*
\section{Formulation of the Rouse model for ring polymer chain}
\label{appendix}

In the Rouse model, the equation of motion for the polymer chain
composed of $N$ beads is given by the following
Langevin equation:
\begin{equation}
	\zeta\frac{d\bm{r}_n}{dt} = -k(2\bm{r}_n - \bm{r}_{n-1} - \bm{r}_{n+1}) + \bm{w}_n(t),
	\label{eq:langevin}	
\end{equation}
where $\bm{r}_n$ represents the coordinates of the $n$-th bead for
$n=1,2,3,\cdots,N$ and 
$\zeta$ denotes the effective hydrodynamic friction coefficient.
Furthermore, two successive beads are connected by a harmonic spring with the
modulus $k$.
Here, the random force $\bm{w}_n$ acting on the bead is related to the temperature $T$ and
friction coefficient $\zeta$ by obeying the fluctuation-dissipation theorem:
\begin{equation}
\langle \bm{w}_n(t) \cdot \bm{w}_m(t')\rangle = 6k_\mathrm{B}T \zeta \delta_{nm}\delta(t-t').
\end{equation}
According to the statistical description for the freely-jointed chain model, the
spring constant $k$ is equal to $3k_\mathrm{B}T/b^2$ with 
the mean square distance $b^2$ between two beads.
Note that the periodic boundary conditions
\begin{equation}
\bm{r}_0=\bm{r}_N,\quad
\bm{r}_{N+1}=\bm{r}_1
\end{equation}
should be imposed on the ring polymer chain.
If we define two $N\times 3$ matrices, 
$\bm{R}=(\bm{r}_1,\bm{r}_2,\bm{r}_3,\cdots,\bm{r}_N)^\mathrm{T}$ and 
$\bm{W}=\zeta^{-1}(\bm{w}_1,\bm{w}_2,\bm{w}_3,\cdots,\bm{w}_N)^\mathrm{T}$
(the superscript T denotes the transpose), 
Eq.(\ref{eq:langevin}) can be expressed as
\begin{equation}
	\frac{d\bm{R}}{dt} = -\frac{k}{\zeta}\bm{A}
	\bm{R} + \bm{W},
\label{eq:langevin2}	
\end{equation}
with the $N\times N$ matrix $\bm{A}$:
\begin{equation}
\bm{A}=
	\begin{pmatrix}
		2 	    &	-1 	& 	0	& 	\cdots	&   0   &   0	& 	-1 	    \\
		-1 	    &	2 	& 	-1 	& 	\cdots	&   0   &   0	&	0 	    \\
		0 	    &	-1	&	2	&	\cdots 	&   0   &   0	&	0 	    \\
		0 	    &	0	&	-1	&	\cdots	&   0   &   0	&	0 	    \\
		\vdots 	&		&		&	\ddots	&       &       &	\vdots 	\\
		0 	    &	0	&	0	&	\cdots	&   -1  &   2	&	-1 	    \\
		-1	    &	0	&	0	&	\cdots	&   0   &   -1	&	2	    \\
	\end{pmatrix}.
 \end{equation}
Equation~(\ref{eq:langevin2}) can be solved by the diagonalization of the matrix $\bm{A}$.
The eigenvalue $\lambda$ equation is given as
\begin{equation}
	(\bm{A}-\lambda\bm{E})\bm{F} = 0,
	\label{eq:eigen}
\end{equation}
with the eigenvector $\bm{F}=(f_1, f_2, f_3,\cdots,f_N)^\mathrm{T}$ and the unit matrix $\bm{E}$.
If the function form of $f_n$ is assumed to be
\begin{equation}
	f_n=z^n,
	\label{eq:eigenfunc_katei}
\end{equation}
with the complex number $z$, 
Eq.~(\ref{eq:eigen}) reduces to the following multiple linear equations:
\begin{gather}
 	(2-\lambda)z -z^2 - z^N =0,\label{eq:1}\\
\quad \quad \vdots \quad \quad \quad \nonumber\\
	-z^{n-1}+ (2-\lambda)z^{n} - z^{n+1} =0,\label{eq:n}\\
\quad \quad \vdots \quad \quad \quad \nonumber\\
	-z - z^{N-1} + (2-\lambda)z^N=0.\label{eq:N}
\end{gather}
From Eq.~(\ref{eq:n}), the characteristic equation
\begin{equation}
-1+(2-\lambda) z - z^2=0,
\end{equation}
is obtained.
The two roots are denoted as $z_1$ and $z_2$, then 
\begin{equation}
z_1+z_2=2-\lambda,\quad
z_1z_2 = 1.
\end{equation}
Furthermore, the function form of $z$ is assumed to be 
\begin{equation}
z_1=e^{i\theta},\quad
z_2=e^{-i\theta}
\end{equation}
such that $z_1z_2 =1$
with the imaginary unit $i$ and an arbitrary argument
$\theta$ in the complex plane.
We obtain the identity:
\begin{equation}
e^{iN\theta} = 1
\label{eq:argument}
\end{equation}
to satisfy Eqs.~(\ref{eq:1}), (\ref{eq:n}), and~(\ref{eq:N}) in a consistent manner.
The argument $\theta$ should be 
\begin{equation}
\theta = \frac{2\pi p}{N},
\label{eq:theta}
\end{equation}
where $p$ denotes the Rouse mode index with $p=0$, 1, 2, $\cdots$, $N-1$.
Thus, the eigenvalue of
the mode $p$ is obtained as
\begin{equation}
	\lambda_p=2-(z_1+z_2)= 2\left(1-\cos\left(\frac{2\pi p}{N}\right)\right)=4\sin^2\left(\frac{\pi p}{N}\right).
	\label{eq:eigenval}
\end{equation}
Note that $\lambda_p=\lambda_{N-p}$.
Accordingly, the Rouse modes are symmetric with respect to the
reflection at $p = N/2$ and the two modes of $p = n$ and $p = N-n$  are
degenerate for ring polymers.

The general solution for the element of the eigenvector $\bm{F}$ can be given by
\begin{align}
f_{n,p}		= Ae^{in\theta} + A^* e^{-in\theta},
\label{eq:eigenfunc}
\end{align}
with a complex constant $A$.
Note that Eq.~(\ref{eq:eigenfunc}) ensures $f_{n,p}=f_{n,p}^*$, where the superscript * denotes the complex conjugate.
The orthogonal condition for $f_{n, p}$ is given by 
\begin{equation}
\sum_{n=1}^N f_{n,p} f_{n,q}^* = \delta_{p,q}.
\label{eq:orthogonal}
\end{equation}
The l.h.s of Eq.~(\ref{eq:orthogonal}) can be expressed as
\begin{multline}
\sum_{n=1}^N (A e^{i2\pi n p/ N} + A^* e^{-i2\pi n p/ N}) \times 
(A^* e^{-i2\pi n q/ N} + A e^{i 2\pi n q/ N})\\
=\sum_{n=1}^N \left(A^2 e^{i 2  \pi n (p+q)/N}
 + AA^* e^{i2  \pi n (p-q)/ N}\right.\\
 \left. + A^*A e^{-i 2  \pi n (p-q)/ N}
+(A^*)^2 e^{-i 2  \pi n (p+q)/N}\right).
\label{eq:orthogonal2}
\end{multline}
To obtain the condition for determining $A$, we assume the special case $p+q=N$
($p\ne q$); 
then, Eq.~(\ref{eq:orthogonal2}) further reduces to 
\begin{align}
\sum_{i=1}^N (A^2 + (A^*)^2),
\end{align}
where $\sum_{n=1}^N e^{i2 n (p-q)/N} = 0$ is used.
Thus, the first relationship $A^2 + (A^*)^2=0$ is obtained from the orthogonal condition, Eq.~(\ref{eq:orthogonal}).
Furthermore, the normalization condition for $f_{n, p}$ is given by 
\begin{align}
\sum^N_{n=1}f_{n,p}f^*_{n,p} 	= 1, 
\end{align}
which can be expressed at $p \ne 0$ or $N/2$ as
\begin{multline}
\sum_{n=1}^N \left(
A^2 e^{i 4\pi n p / N} + 2AA^* + (A^*)^2 e^{-i 4 \pi n p/N}
\right)
= \sum_{i=1}^N 2AA^* = 1,
\end{multline}
and the second relationship $AA^* = 1/(2N)$ is obtained.
We again use $\sum_{i=1}^N e^{i4 \pi np / N}=0$ in the cases of $p\ne 0$
and $p\ne N/2$.
Note that $AA^* = 1/(2N)$ is also obtained in the 
two cases $p=0$ and $p=N/2$ according to $A^2+(A^*)^2=0$.
From the two relationships, the complex constant $A$ can be determined, and its
expression is chosen from four candidates:
$A=(1+i)/(2\sqrt{N})$,  $(-1-i)/(2\sqrt{N})$, $(-1+i)/(2\sqrt{N})$, and $(1-i)/(2\sqrt{N})$.
The functional form of $f_{n,p}$ is then determined as
\begin{equation}
f_{n,p} = \sqrt{\frac{1}{N}}\left[\cos\left(\frac{2\pi np}{N}\right) +
			     \sin\left(\frac{2\pi np}{N}\right)\right], 
\label{eq:rouse_mode}
\end{equation}
and Eq.~(\ref{eq:rouse_mode}) satisfies Eq.~(\ref{eq:orthogonal}).
Note that a different expression for $f_{n,p}$ is described and 
utilized in
the path integral molecular dynamics.~\cite{Ceriotti:2010hm}

Here, we define the block matrix composed of the orthonormal
eigenvectors, 
$\bm{U}= (\bm{U}_0, \bm{U}_1, \cdots, \bm{U}_{N-1})$, with
$\bm{U}_p
= (f_{1,p}, f_{2,p}, \cdots, f_{N,p})^\mathrm{T}$, which diagonalizes
the matrix $\bm{A}$ as 
\begin{equation}
\bm{U}^\mathrm{T} \bm{A} \bm{U} = 
\begin{pmatrix}
\lambda_0 & 0 &   \cdots & 0\\
0 & \lambda_1 &  \cdots &  0\\
\vdots & \vdots       & \ddots & \vdots \\
0 &  0 & \cdots & \lambda_{N-1}
\end{pmatrix}.
\end{equation}
The normal coordinates are finally described as 
\begin{align}
\bm{X} =\bm{U}^\mathrm{T}\bm{R}, 
\end{align}
with the element 
\begin{align}
\bm{X}_p = 
 \sqrt{\frac{1}{N}}\sum_{n=1}^{N}
\bm{r}_n(t)\left[\cos\left(\frac{2\pi np}{N}\right) + \sin\left(\frac{2\pi np}{N}\right)\right],
\end{align}
for the ring polymer chain.

From Eq.~(\ref{eq:langevin2}),
the normal coordinates of mode $p$ obeys the following equation: 
\begin{equation}
 	\frac{d\bm{X}_p}{dt} = -\frac{k}{\zeta}\lambda_p \bm{X}_p+\bm{W}_p,
	\label{eq:rouse}
\end{equation}
where 
$\bm{W}_p = \bm{U}^\mathrm{T}\bm{W}$ is the random force, which satisfies 
\begin{equation}
\langle \bm{W}_p(t) \cdot \bm{W}_q(t')\rangle = 6k_\mathrm{B}T \zeta^{-1} \delta_{p,q}\delta(t-t').
\label{eq:wpwp}
\end{equation}
The formal solution of Eq.~(\ref{eq:rouse}) is given by 
\begin{equation}
\bm{X}_p(t)=\bm{X}_p(0)\exp(-t/\tau_p) + \int^t_0dt'  \bm{W}_p(t')\exp(-(t-t')/\tau_p),
\end{equation}
where
\begin{equation}
\tau_p=\frac{\zeta}{k\lambda_p} = \frac{\zeta}{4k\sin^2\left(\frac{\pi p}{N}\right)}
\end{equation}
represents the Rouse relaxation time.
The autocorrelation function of $\bm{X}_p(t)$ is generally described by 
\begin{equation}
\langle \bm{X}_p(t)\cdot \bm{X}_p(0) \rangle =
 \frac{3k_\mathrm{B}T}{k\lambda_p}\exp(-t/\tau_p).
\label{eq:autocorrelation}
\end{equation}
The static correlation of the Rouse mode is expressed as 
\begin{equation}
\langle \bm{X}_p(0)^2 \rangle = \frac{3k_\mathrm{B}T}{k\lambda_p}=\frac{b^2}{4\sin^2\left(\frac{\pi
 p}{N}\right)}
\end{equation}
from the the initial value of Eq.~(\ref{eq:autocorrelation}).

\begin{acknowledgments}
This work was supported by JSPS KAKENHI Grant Numbers:
JP18H01188 (K.K.), JP20H05221 (K.K.), and JP19H04206 (N.M.).
This work was also partially supported 
by the Fugaku Supercomputing Project (No.~\mbox{JPMXP1020200308}) and the Elements Strategy
Initiative for Catalysts and Batteries (No.~\mbox{JPMXP0112101003}) from the
Ministry of Education, Culture, Sports, Science, and Technology.
The numerical calculations were performed at Research Center of
Computational Science, Okazaki Research Facilities, National Institutes
of Natural Sciences, Japan.
\end{acknowledgments}


\section*{Conflicts of Interest}
The authors have no conflicts to disclose.

\section*{data availability}
The data that support the findings of this study are available from the
corresponding authors upon reasonable request.

%

\end{document}